\documentclass[epj,final]{svjour}

\usepackage{graphicx}
\usepackage{color}

\newcommand{\bea}{\begin{eqnarray}}
\newcommand{\eea}{\end{eqnarray}}
\newcommand{\beq}{\begin{equation}}
\newcommand{\eeq}{\end{equation}}

\newcommand{\tref}[1]{(\ref{#1})}

\newcommand{\tsemat}[1]{{\mathbf{\textsf{#1}}}}

\newcommand{\Amat}{\tsemat{A}}
\newcommand{\Bmat}{\tsemat{B}}
\newcommand{\Btildemat}{\tilde{\Bmat}}

\newcommand{\Cmat}{\tsemat{C}}

\newcommand{\Emat}{\tsemat{E}}

\newcommand{\unitmat}{\hbox{\textsf{1}\kern-.25em{\textsf{I}}}}

\newcommand{\Btilde}{\tilde{B}}

\newcommand{\str}{s}

\newcommand{\strout}{s^{(\mathrm{out})}}

\begin{document}

\title{Line Graphs of Weighted Networks for Overlapping Communities}

\author{T.S.\ Evans\inst{1,2}\thanks{\email{t.evans@imperial.ac.uk}} \and R.\ Lambiotte\inst{1}\thanks{\email{r.lambiotte@imperial.ac.uk}}
}                     
%
%
\institute{Institute for Mathematical Sciences, Imperial College London, SW7 2PG London, UK
 \and
Theoretical Physics, Imperial College London, SW7 2AZ, U.K.}
%
\date{9th June 2010}
%
\abstract{In this paper, we develop the idea to partition the edges of a weighted graph in order to uncover overlapping communities of its nodes. Our approach is based on the construction of different types of weighted line graphs, i.e.\ graphs whose nodes are the links of the original graph, that encapsulate differently the relations between the edges. Weighted line graphs are argued to provide an alternative, valuable representation of the system's topology, and are shown to have important applications in community detection, as the usual node partition of a line graph naturally leads to an edge partition of the original graph. This identification allows us to use traditional partitioning methods in order to address the long-standing problem of the detection of overlapping communities. We apply it to the analysis of different social and geographical networks.
} 

\maketitle



\section{Introduction}\label{sintro}

In the last decade, the interdisciplinary field of complex networks has led to the development of universal tools in order to characterise and model systems as diverse as information, biological or social networks \cite{review}. Many studies focus on the properties of the vertices, e.g.\ studying their degree distribution or ranking them by some measure.  However graphs are both a set of vertices and a set of relationships between vertices --- the edges.  It is therefore useful sometimes to look at a network from the view point of the edges.  We do this by defining `weighted line graphs' for any type of graph, extending our original work on weighted line graphs for simple graphs \cite{EL09}.  Our weighted line graphs are topologically equivalent to the standard line graph of the literature \cite{Whitney,HN60,RB78}.  However the weights we define play a crucial role in avoiding a bias inherent in unweighted line graphs towards high degree vertices in the original graph. Our work can be seen as providing a general framework to shift our view from a vertex centric one to an edge centric viewpoint.

We illustrate our ideas in the context of community detection \cite{F09,MOM09,LF09,GL08}. When dealing with complex networks one crucial step is the identification of communities or modules, some sort of highly connected subgraphs. It has been shown that many systems of interest are organised in a modular way and that these topological modules usually correspond to functional sub-units. In a large number of situations, these building blocks themselves may be modular, in which case the network is said to be hierarchical. Modularity at different scales has long been argued to be a universal property of complex systems because of the crucial evolutionary advantage it confers, by providing stable intermediate forms (modules) and thereby improving the system's adaptability \cite{simon}. Multi-scale modularity is also associated to a separation of time scales for the dynamics taking place on the graph \cite{syn1,rosvall,DYB08,LDB08}, which is essential in order to ensure the persistence of diversity in the system \cite{lambi}.

The fundamental idea behind most community detection methods is to partition the nodes of the network into modules. By doing so, each node is therefore assigned to one single module. However a vertex partition has the disadvantage of being incompatible with the existence of overlapping communities, i.e.\ situations where nodes belong to several communities.  This overlap is known to be present at the interface between modules, but can also be pervasive in the whole network \cite{ABL09}. This is the case in many social networks where individuals typically belong to several communities defined by their type of interaction, e.g. work, sport buddy, family, etc, but also in biological networks where proteins may belong to several functional categories. In those situations where the interface between the communities occurs throughout the system, a partition of the nodes is questionable as it imposes undesired constraints on the community detection problem. There are many different approaches to finding overlapping communities  (for example see \cite{BGM05,PDF,LLY06,G08b,NMCM08,EL09,LFK09,ABL09,SSA09,WQWZ09,P09,SCG09,SCZ10}). A popular choice is $k$-clique percolation, which consists in looking for connected components of cliques of size $k$ \cite{PDF}. However, this approach has several disadvantages as its outcome strongly depends on the sparsity of the network, it has a single integer parameter with which to set the scale of communities found, it is not easily implementable for weighted networks, and is not applicable to multi-scale networks.  For instance it fails on one of the classic tests for community detection algorithms, the Karate club graph of Zachary \cite{Z77}.

Our approach is based on the observation that, even if nodes may belong to multiple groups, links often correspond to one particular type of interaction. For instance, in the case of social networks the connection between two people is usually for one dominant reason (work, sport interest or family). In contrast to nodes, links therefore typically belong to one single module. In order to exploit this observation, we define communities as partitions of links rather than of nodes.  The edges incident at a single node may belong to several modules and in this sense, nodes can be members of several communities. This change of perspective has several advantages. First, it is a very simple idea.  It is perhaps surprising that we have few other attempts to define simple edge partitions. Secondly, it is a very general, flexible framework. We simply apply standard vertex partitioning to the weighted line graphs defined below. Thirdly, link partitions naturally produce overlapping communities while uncovering a multi-scale, hierarchical organisation. Indeed, the different levels of a dendrogram correspond to partitions whose communities are nested in each other. Uncovering edge partitions at different scales is therefore capable of revealing the hierarchical, overlapping structure of a network. Finally, our approach can easily be generalised in order to analyse weighted and/or directed networks.

This article is organised as follows.  First we recall from \cite{EL09} how to construct various useful types of line graphs of simple graphs, and expose the central ideas of our approach. In the section 3, we show how to generalise the method to weighted graphs and how to overcome the complications which arise in this case.  In section 4, we show some examples of how our methods work in the context of community detection. In section 5, we discuss possible generalisations of our work to the case of multigraphs and directed graphs. In section 6, finally, we summarise our findings and conclude.

\section{Simple Graphs $G$}\label{suwudg}

\subsection*{Overview}

In our approach we find it useful to start from the representation of a network $G$ in terms of its
incidence matrix $\Bmat$. Suppose our original simple graph $G$ has $N$ vertices, which we will label with mid-alphabet Latin characters $i,j, \ldots$, and $L$ edges which we label with early Greek alphabet characters $\alpha, \beta, \ldots$. We define the incidence matrix\footnote{This can be considered to be the adjacency matrix of a bipartite graph.  This graph is a special case of what is known as incidence graph --- the incidence of a set of lines with a set of points in a Euclidean space of finite dimension.} of a simple graph $G$, $\Bmat(G)$, such that $B_{i \alpha}$ is $1$ if link $\alpha$ is related to node $i$, otherwise they are $0$. This contains all the information about the graph $G$.  For instance the adjacency matrix $\Amat$ of the graph $G$ is given by
\begin{equation}
A_{ij}
 = \sum_{\alpha} B_{i \alpha} B_{j \alpha} (1- \delta_{ij} )\, .
\label{adjdef}
\end{equation}
Thus the degree of a vertex is $k_i=\sum_j A_{ij}$.

We will use the concept of random walkers on graphs to motivate our choice of weights in our weighted line graphs.
In terms of the vertices of $G$, the usual random walk process is defined such that at each step the walkers move from their current vertex to a neighbouring one chosen with equal probability.  Thus the density of random walkers on node $i$ at step $n$ is $p_{i;n}$ where
\begin{equation}
\label{discrete} p_{i;n+1} = \sum_{j} \frac{A_{ij}}{k_j} \,
p_{j;n} \, .
\end{equation}

As we look at community detection on our weighted line graphs, it is useful to note here that the widely-used Newman-Girvan ``modularity'' $Q$ \cite{GN04} can be interpreted in this dynamical context \cite{DYB08,LDB08}.  The best vertex partition of the graph is often found by maximising Newman-Girvan modularity which measures if there are more edges within communities than would be expected on the basis of chance. The quality function maximised is the modularity $Q$ where\footnote{We also note that communities at different scales can be found by introducing a resolution parameter in the definition of modularity  \cite{RB06,AFG08}.}
\begin{eqnarray}
Q(\Amat) = \sum_{C \in \mathcal{P}} \sum_{i,j \in C}
\left[ \frac{A_{ij}}{k_j}\pi_j - \pi_i \pi_j \right] \,.
\label{modAgendef}
\end{eqnarray}
Here $\pi_j= \lim_{n \rightarrow \infty} p_{i;n}$ is the long time distribution of random walkers, which is well-defined and unique if the dynamics is ergodic.  For simple graph it is given by $\pi_i =k_i/W$, where $W=\sum_{i,j}A_{ij}$, under quite general circumstances \cite{Chung}.  The indices $i$ and $j$ run over the nodes in community $C$ while $C$ is taken through the different communities of the vertex partition $\mathcal{P}$. Modularity is therefore equivalent to the probability of a random walker to remain in the same community over two successive time steps, minus the probability for independent walkers to be in those communities at those times. A partition which gives a large value of $Q$ is usually taken to be a good community structure for the graph $G$.

\subsection*{Random walk on the edges and weighted line graphs}

\begin{figure}
\begin{center}
\includegraphics[width=0.33\textwidth]{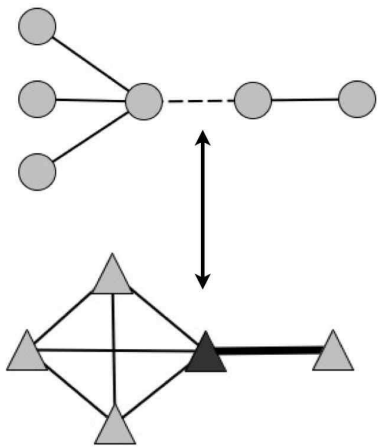}
\end{center}
\caption{The weighted line graph transformation emphasises the role of edges in the network while properly accounting for the degree heterogeneity present in the network. Each link in the original simple graph (top) corresponds to a node in the line graph (bottom) while nodes transform into weighted cliques. The ``Link-Node-Link random walk'' on the original graph, as defined in the text, is equivalent to an unbiased random walk on the nodes of the weighted line graph. In this illustration, the width of the links is proportional to their weight and the dotted link is transformed into the darkened node.}\label{frandwalks}
\end{figure}

Our desire to move from a vertex centric viewpoint to one focussed on edges, suggest that we consider  random walkers moving from edge to edge.  On a simple graph, each step of such a walk has two characteristic quantities to consider, the degree of the vertices at each end $k_i$ and $k_j$.  This leads naturally to two different processes \cite{EL09}:
\begin{itemize}
\item a random walk where the walkers can jump to all available edges with equal probability, namely $1/(k_i + k_j - 2)$. When $k_i \neq k_j$, the walker goes with a different probability through $i$ or $j$, and we therefore call this process a ``link-link random walk'' ;
\item a ``link-node-link random walk'', where a walker first jumps with equal probability to one of the two nodes to which it is attached, say $i$.  It then moves to a new link incident at $i$, again choosing with equal probability from those available.  Thus with probability $1/( 2 (k_i -1))$ it ends on one of the links leaving $i$ and with probability $1/( 2 (k_j - 1))$ it finishes on a new link leaving $j$.  As this process is not defined for vertices of degree one we ignore such vertices and so the walker will always jump to the other vertex.
\end{itemize}

The simplest way to shift the focus from vertices to edges is to construct the other product from the rectangular incidence matrix $\Bmat$.  Thus we define the line graph $L(G)$ through its $L \times L$
adjacency matrix $\Cmat$:
\begin{equation}
C_{\alpha \beta} = \sum_{i} B_{i \alpha} B_{i \beta} (1-
\delta_{\alpha \beta}). \label{adjCuwud}
\end{equation}
The line graph is a well known construction \cite{Whitney,HN60,RB78} that almost perfectly encodes the topological properties of the original graph. The structure of $G$ can be recovered completely from its line graph $L(G)$, for almost any graph except for a triangle or a star network of four nodes \cite{Whitney}. The vertices of the line graph are in one-to-one correspondence with the edges of the original graph $G$, except for the edges of leaves (i.e.\ edges which end in a degree one vertex). A vertex in the original graph of degree $k$ is mapped into $k(k-1)/2$ edges of the line graph.

If we now perform the usual vertex random walk on the vertices of the line graph $C(G)$ we see that this corresponds to
\begin{equation}
p_{\alpha;n+1} = \sum_{\beta} \frac{C_{\alpha \beta}}{k_\beta}
\, p_{\beta;n}.
\end{equation}
where $k_\alpha=\sum_\beta C_{\alpha
\beta}=(k_i + k_j -2)$ and $i$ and $j$ are the
vertices at the end of edge $\alpha$ in the original graph $G$.  Consequently, we observe that the usual random walk on the vertices of this line graph $C(G)$ corresponds to a ``link-link random walk'' on the edges $\alpha$ of $G$.  It is interesting to note that this type of line graph has found many applications in recent years, see for instance \cite{H99,NUYKA04,PEO04,NYGKA05,ZLNZ06,AS06,MSCB09,MMK10}.
However, its big drawback is that each vertex $i$ in the original graph $G$ contributes $k(k-1)/2$ edges to $C(G)$ even though its importance in the original graph could be estimated to be just $k$.  That is the large degree vertices, the hubs, are given too much prominence in the line graph \cite{EL09,ABL09}.

The solution suggested in \cite{EL09} is to define a new type of line graph, the weighted line graph
$D(G)$ with adjacency matrix
\begin{equation}
D_{\alpha \beta} = \sum_{i, k_i >1} \frac{B_{i \alpha} B_{i
\beta}}{k_i -1} (1- \delta_{\alpha \beta}). \label{adjDuwud}
\end{equation}
In the context of projecting
bipartite networks this is a well known weighting \cite{project2}.  If we consider the usual vertex random walk on this line graph $D(G)$, so
\begin{equation}
\label{rw2}
p_{\alpha;n+1} = \sum_{\beta}
\frac{D_{\alpha \beta}}{k_\beta} \, p_{\beta;n}
\end{equation}
then we see that this is equivalent to a link-node-link random walk on the original graph $G$, see Fig \ref{frandwalks}B.

\subsection*{Central idea}

At the heart of our approach is the construction of a line graph in order to represent the system from an edge centric viewpoint. As we have shown in the previous section, there exist different ways to project the incidence matrix onto a line graph, and each projection is associated to a different dynamics taking place on the edges, i.e., to a different interpretation of what the relations between edges are. As we will see in the next section, the number of ways to construct a line graph, when the original graph is weighted, is even larger. The selection of a sensible projection is therefore an essential ingredient, which may in principle depend on the system under scrutiny but should in any case avoid biasing the representation of the network, for instance by giving too much importance to certain nodes. This is the reason why $D(G)$ is preferred to $C(G)$ when analysing simple graphs \cite{EL09}.

\section{Undirected Weighted Graphs $G$}

Suppose now we have an undirected but weighted graph $G$. The incidence matrix may be defined as before to be $B_{i \alpha} = 1$ if edge $\alpha$ is incident to vertex $i$ with all other entries in these rectangular incident matrices are zero.  To record the weights of the edges it is useful to define a second weighted incidence matrix $\Btildemat$ as
\beq
 \Btilde_{\alpha j } =  w_\alpha
 \label{incBwud}
\eeq
where edge $\alpha$ is incident on vertex j and has weight $w_\alpha$. Each vertex then has degree $k_i$ and strength $\str_i$ given by
\beq
 k_i = \sum_{j}\theta(A_{ij}) = \sum_\alpha B_{i \alpha} \, ,
 \;
 \str_j = \sum_{i}A_{ij}  = \sum_\alpha \Btilde_{\alpha j} \, .
 \label{strdefUD}
\eeq
The adjacency matrix of the original graph $G$ is then
\begin{equation}
 A_{ij}
  = \sum_{\alpha = (i,j)} B_{i \alpha} \Btilde_{j \alpha}
  = \sum_{\alpha = (i,j)} w_\alpha \, ,
\label{adjAudw}
\end{equation}
where $\alpha = (i,j)$ indicates that then sum is taken over all edges from vertex $j$ to $i$. This matrix is symmetric as required.

\begin{figure}
\begin{center}
\includegraphics[width=0.33\textwidth]{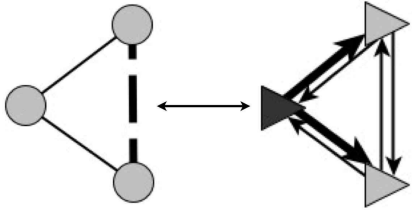}
\end{center}
\caption{When applied to the weighted but undirected network on the left (width of the links is proportional to their weight in this illustration), the weighted line graph transformation leads to the weighted and directed network shown on the right. In this example, the dotted link is transformed into the darkened node.}\label{frandwalks2}
\end{figure}

If we wish to use the weight information of $G$, the logical generalisation of the definitions for $\Cmat$ for unweighted graphs $G$ of \cite{EL09} is as follows\footnote{If we ignore the weights completely then we get a line graph which is the traditional unweighted one, $L(G)$. This would be defined using only $\Bmat$ as ${L}_{\alpha \beta}
 = \sum_i B_{\alpha i} B_{i \beta} (1 - \delta_{\alpha\beta})$. This representation only records the
topological information of the original graph.}:
\begin{equation}
 C_{\alpha \beta}
 = \sum_i \Btilde_{\alpha i} B_{i \beta} (1 - \delta_{\alpha\beta})
 \label{adjCudw}
\end{equation}
This definition for the adjacency matrix of a line graph mimics our construction of the adjacency matrix $\Amat$ of the graph $G$ in \tref{adjAudw} which also used both $\Bmat$ and $\Btildemat$. However, even if the original graph $G$ is undirected, this adjacency matrix is not symmetric, i.e., the line graph $C(G)$ is directed.
If we think in terms of random walks from edge $\beta$ to vertex $i$ and then to edge $\alpha$ then it is natural that the edge weights are linked to the stubs leaving vertex $i$, hence the use of $\Btildemat$ in \tref{adjCudw}. The probability of moving to an adjacent edge is proportional to the target edge's weight $w_\alpha$ but is independent of the current edge's weight $w_\beta$.

The problem with the definition of $\Cmat$ in \tref{adjCudw} is that even though it involves the weights of the edges through $\Btildemat$, a vertex of strength $s$ in graph $G$ is going to contribute $O(k \str)$ to the total weight of these line graphs, which seems like over counting. High degree, high strength vertices are too prominent.  The solution is to reduce the weight of assigned to each link in the weighted line graph by $O(s^{-1})$. Thus we consider the adjacency matrix
\begin{equation}
 E_{\alpha \beta} =
 \sum_{i, k_i>1}
 \frac{\Btilde_{\alpha i} }{\str_i -w_\beta} B_{i \beta}
 (1- \delta_{\alpha \beta}) \, .
 \label{adjEudw}
\end{equation}
This is also a more natural definition when we consider the dynamics of a random walker moving from edge $\beta$ to vertex $i$ and then to edge $\alpha$.  The first step is to each end of the edge $\beta$ with equal probability ($B_{i\beta}$ term) while the latter step to arrive at edge $\alpha$ is taken in proportion to the weights of the edges at $i$ ($\Btilde_{\alpha i}$ term). There exist many other ways to project the incidence graph $B(G)$ onto a weighted line graph\footnote{Other interesting generalisations include $ D_{\alpha \beta} =
 \sum_{i, k_i >1}
 \frac{\Btilde_{\alpha i}  }{k_i -1} B_{i \beta}
 (1- \delta_{\alpha \beta})$
 and $ F_{\alpha \beta} =
 \sum_{i, k_i >1}
 \frac{\Btilde_{\alpha i} }{(\str_i -w_\beta)(k_i-1)} B_{i \beta}
 (1- \delta_{\alpha \beta})$.}
but this definition is the one which preserves the dynamics of random walkers.  The dynamics of random walkers is important in many contexts of graph theory, such as in the PageRank algorithm or in the context of Newman-Girvan modularity $Q$ \tref{modAgendef} as noted above.

When the original graph $G$ is unweighted and undirected then this weighted line graph $E(G)$ reduces to the weighted line graph described in \cite{EL09}.  However if the original graph $G$ is weighted then the weighted line graph $E(G)$ will be both directed and weighted. One special case is when the original graph $G$ is ergodic in which case so is this weighted line graph $E(G)$.

\section{Applications}

Once the projection from a weighted graph $G$ to the weighted line graph $E(G)$ \tref{adjEudw} to has been made, it is possible to use any vertex metric on the line graph in order to characterise the structure of the edge sin the original graph. It is for instance possible to look at the centrality or the clustering coefficient of the nodes of the line graph in order to uncover the role of the original edges. A study of the degree distribution in the line graph is sensitive to degree-degree correlations of neighbouring vertices in the original graph.

Here though we will focus on the vertex partition of the weighted line graph $E(G)$ \tref{adjEudw} in order to produce an edge partition of the original graph $G$.  In principle, any vertex partitioning scheme can be used. However since optimisation of modularity is related to the behaviour of random walkers on a graph and our construction of $E(G)$ preserves the dynamics of random walkers, it makes sense to apply the modularity optimisation approach to find the partitions of the weighted line graph $E(G)$ \tref{adjEudw}.  So we will search for maxima of
\begin{equation}
Q(\Emat) = \sum_{C \in \mathcal{P}} \sum_{\alpha, \beta \in C} \left[ \frac{E_{\alpha \beta}}{\strout_\beta}\pi_\beta
      - \pi_\alpha \pi_\beta \right],
      \label{modQE}
\end{equation}
where the out-strength is $\strout_\beta=\sum_\beta E_{\alpha \beta}$.   The vector $\pi_\beta$ is the dominant eigenvector of the transition matrix $(E_{\alpha \beta} /\strout_\beta)$  with eigenvalue one, normalised such that $\sum_\alpha \pi_\alpha =1$. Let us emphasise that a weighted but undirected graph $G$ produces a weighted line graph $E(G)$ which is also directed, so that the equilibrium walker distribution $\pi_\alpha$ is non-trivial.  This has to be computed first, which we do by using the power method \cite{power}.

Maxima of $Q(\Emat)$  \tref{modQE} can rarely be found exactly but there are many good approximate algorithms.  For our own convenience we use the Louvain algorithm of \cite{Blondel} to find a partition of the vertices of $E(G)$ which gives a large value of modularity $Q(\Emat)$.

\subsection*{Literary Characters Coappearance}

\begin{figure}[thbp]
\includegraphics[width=\columnwidth]{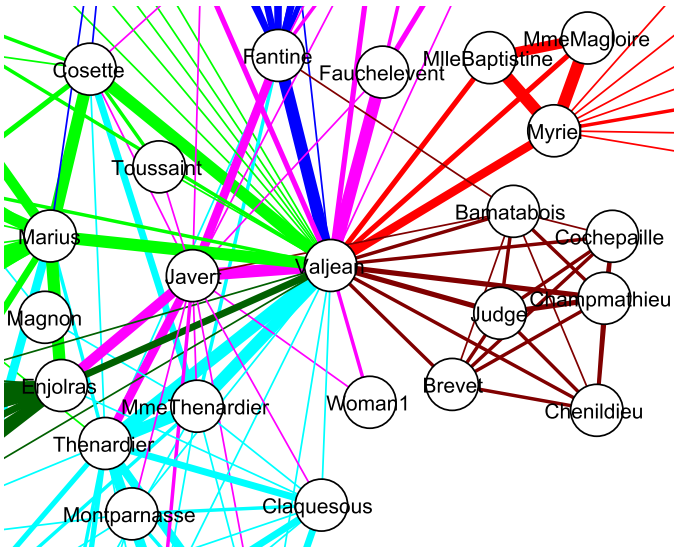}
\caption{Part of the graph of characters in Les Miserables, centred on the main character Valjean. Characters are linked by an edge if they appear in the same scene and the weight is equal to the number of chapters in which they both appear \cite{K93a}. The edge colours reflect a partition which produces an approximate maximal value of $Q(E)$.  This method allows vertices to be a member of many communities, appropriate for many characters such as Valjean shown here.}\label{flesmis}
\end{figure}

Our first example of a weighted graph is based on the appearances of characters in the same chapter of Les Miserables \cite{K93a}.   The vertices are different characters and the weight of edges is the number of chapters in which that pair of characters has appeared together.  The results of performing a vertex partition on the line graph $E(G)$ are shown in Fig.\ref{flesmis}.  The result is generally compatible with the vertex partition found in \cite{GN04} and presumably reflect the natural communities that a narrative structure will produce in many novels and plays.  However the main advantage our edge colouring approach is that characters, especially the main ones, will belong to several communities, as indicated by the different coloured edges.  In particular the main protagonist, the vertex labelled Valjean in Fig.\ref{flesmis}, is connected to all but one community but the strength of his connection to each community varies significantly as Table \ref{tlesmis} shows.

\begin{table}[htbp]
\begin{center}
\begin{tabular}{r|c}
Community  & Valjean Membership  \\ \hline
Myriel     &  7\% \\ \hline
Marius     & 38\% \\ \hline
Fantine    &  6\% \\ \hline
Thenardier & 15\% \\ \hline
Javert     & 22\% \\ \hline
Judge      &  9\% \\ \hline
Enroljas   &  4\%
\end{tabular}
\end{center}
\caption{Table showing the fraction of edge weight incident at the Valjean vertex in the communities found by optimising the  modularity $Q(E)$ of \tref{modQE}. Communities are labelled by the character (other than Valjean) with the largest weight of edges in that community. }
\label{tlesmis}
\end{table}

\subsection*{Clustering Non Negative Matrices}

\begin{figure}[htbp]
\includegraphics[width=\columnwidth]{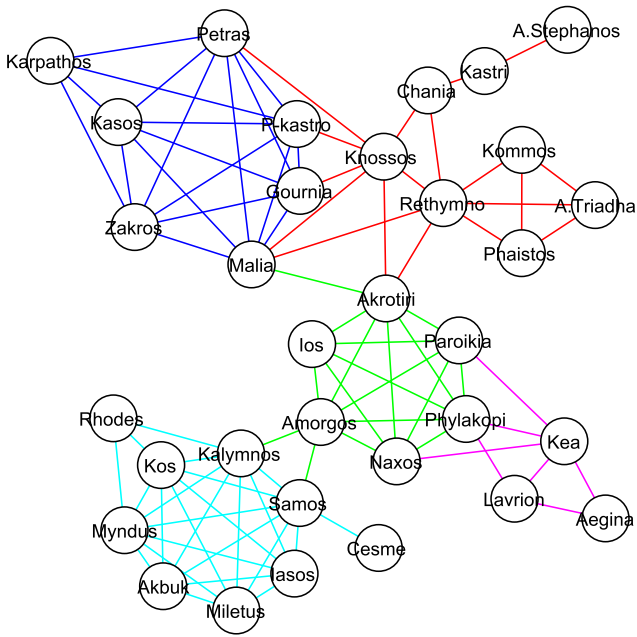}
\caption{The edge partition of a graph of Middle Bronze Age sites in the Aegean. The weight of an edge is $\theta((1+(x)^4)^{-1}-0.220)$ where $d$ is the distance in 100 kilometres between two sites. 100km is roughly the distance one could travel in a day. The distances have been estimated using the shortest route where land travel is weighted by a factor of 3.0 while sea travel is weighted by 1.0 \cite{KER08}.  The threshold of 0.220 is chosen such that 33 of the 34 sites form a connected graph. The edge colours reflect a partition which produces an approximate maximal value of $Q(E)$. }\label{faegean}
\end{figure}

It is common to come across dense matrices with non-negative entries.  One will often be interested in reducing the dimension of the space by looking for clusters of entries which are similar in some sense.  By converting these matrices into a sparse graph, the problem becomes equivalent to the search for communities in networks.

\begin{figure}[htbp]
\includegraphics[width=\columnwidth]{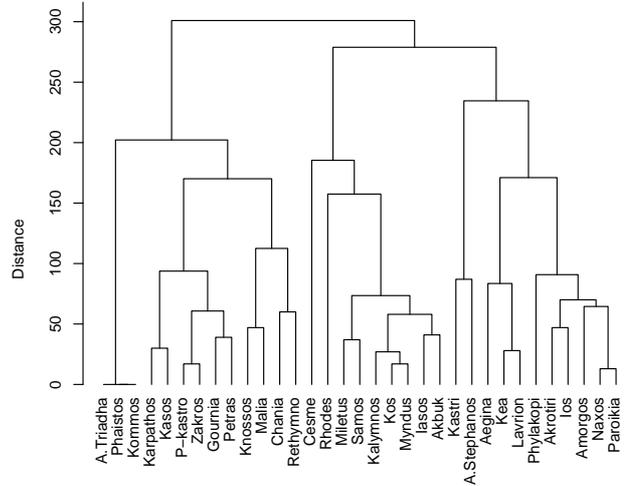}
\caption{A dendrogram derived from the matrix of distances between 33 key sites of the Middle Bronze Age in the Aegean.  The horizontal lines indicate the average distance between the groups of sites indicated by the vertical lines below that horizontal line.}\label{faegeandend}
\end{figure}

We illustrate our approach with an example of geographical separation of sites.  We consider a set of 33 important Middle Bronze Age sites in the Aegean (c. 2000BC-1400BC) taken from \cite{KER08,EKR09}.  In the corresponding graph, the sites are vertices and edges are given a weight which is a monotonically decreasing function of the distance between two sites.  Finally to produce a sparse graph a threshold is used and any edge with weight below this value is removed.   The edge partition of this graph found by optimising the modularity of the line graph $E(G)$ is shown in Fig \ref{faegean}.  This produces five communities: Asia Minor and the Dodecanese (Miletus), the Cyclades (Naxos), Eastern Crete (Palaikastro), Central and Western Crete (Knossos) and a small group centred on Attica (Aegina).  A vertex partition might well uncover similar groups but it would not emphasise that some sites may have a more complex relationship to the main groups.  For instance, Akrotiri on modern Santorini in the Cyclades is part of both the Cycladean and a Cretan community.  This emphasises the role it may have played in the both in expansion of Minoan influence during this era, and in its demise following the destruction of Akrotiri in the eruption of ancient Thera (Santorini is the modern remnant).  Another way to see the usefulness of this type of approach is to compare against a more traditional dendrogram analysis of the distance matrix, such as shown in Fig \ref{faegeandend}.  For instance the special role of Akotriri is not apparent in the dendrogram of Fig \ref{faegeandend}.

\subsection*{Academic Coauthorship}

\begin{figure}[htbp]
\includegraphics[width=\columnwidth]{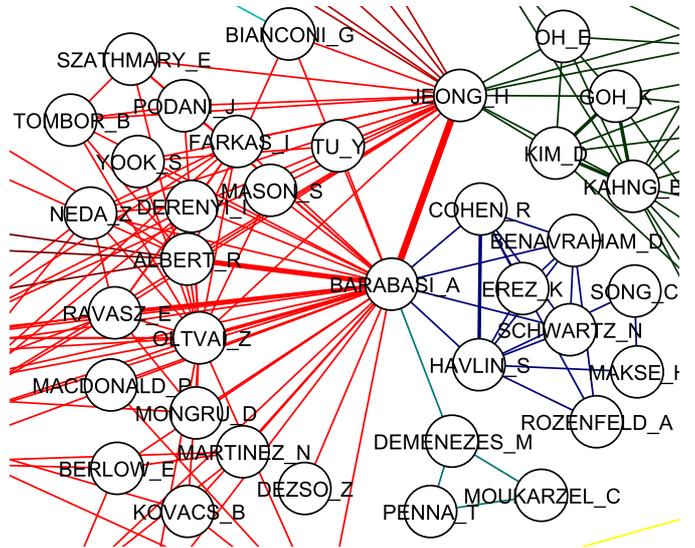}
\caption{Part of the coauthorship network of scientists, as defined by Newman \cite{N06}. Each paper of $k$ authors contributes a weight of $(k-1)^{-1}$ to an edge between each of the $k(k-1)/2$ pairs of collaborators. The edge colours reflect a partition which produces an approximate maximal value of $Q(E)$.}\label{fnetscibar}
\end{figure}

In Fig\ \ref{fnetscibar} we show part of the weighted graph representing the coauthorships of scientists on some network papers, as defined by Newman \cite{N06}.  The edges are partitioned by searching for a large $Q(E)$.  Here we find that some of the most productive scientists are the focus of one community, and they participate in other communities much less often.  The links between these groups are often provided by less prominent researchers, reminding one of the strength of weak links hypothesis of Granovetter \cite{Granovetter1973}.  For instance in Fig \ref{fnetscibar} Barab\'{a}si is the centre of one main community though a few edges incident at the Barab\'{a}si vertex are also part of two other communities.

\section{Possible generalisations}

In this paper, we have focused on line graphs without self-loops. However there are natural alternatives to our definitions which include self-loops in the line graphs \cite{EL09}. Their adjacency matrices take the form $\sum_i \Btilde_{\alpha i} B_{i \beta}/v_i$ where obvious choices for $v_i$ are $1$, the degree $k_i$, the strength $s_i$ or the product $(k_i s_i)$ which are the analogues of $C(G)$ \tref{adjCudw}, $D(G)$, $E(G)$ \tref{adjEudw}, and $F(G)$ respectively. One advantage of these line graphs have over our previous definitions is that all connected vertices are explicitly represented in these graphs.  The presence of self loops corresponds to allowing random walkers to move first to either vertex at the ends of an undirected edge, but then being allowed to come back to finish on the same edge it started from. Whether this type of random walk and these line graphs are a better way of studying the graph $G$ will depend on the context.  Interestingly, in the context of community detection, adding self loops is a technique used to alter the resolution of algorithms \cite{AFG08}.  Thus it may be that for community detection there is little difference in practice if one also alters the number of communities found by an algorithm e.g\ by altering modularity \cite{RB06,AFG08}.

Our formalism can also be generalised to situations when the original graphs $G$ have self-loops or multiple edges between vertices, which has not been considered so far. Indeed, self-loops and multiple edges are correctly encoded in the incidence matrix representation $B(G)$ of \tref{incBwud}.  The presence of self-loops requires some adaptation of our formulae but multigraphs are included without any change. A multigraph representation could have interesting consequences, as it could allow edges to be a member of several different communities.  In this case the original edge is split into several edges whose total weight is equal to that of the original edge.  In social networks this means the relationship between two individuals can be of more than one type, e.g.\ two work colleagues may also share the same hobby.

Finally our results can be generalised to cases where the original graph itself is directed. To do so, we propose to look at the unweighted incidence matrix $\Bmat$ in terms of the incoming edges, that is $B_{i \alpha} = 1$ if edge $\alpha$ goes into vertex $i$.  The weighted incidence matrix $\Btildemat$ would be defined in terms of the source vertex of an edge and its weight, so $ \Btilde_{\alpha j} =  w_\alpha$ if edge $\alpha$ of weight $w_\alpha$ is leaving vertex $j$.  The adjacency matrix of $G$ is then
\begin{equation}
A_{ij} = \sum_{\alpha} B_{i \alpha} \Btilde_{\alpha j},
\end{equation}
 while the adjacency matrices of the line graphs are given by
\begin{equation}
 \sum_{i, v_i >0} \frac{\Btilde_{\alpha i} }{v_i } B_{i \beta},
 \end{equation}
where $v_i$ can be $1$ for $C(G)$, $k_i= \sum_\alpha \theta(\Btilde_{\alpha i})$ for $D(G)$, $\str_i = \sum_\alpha \Btilde_{\alpha i}$ for $E(G)$, or $(k_is_i)$ for $F(G)$. It is interesting to note that a random walker performing a link-node-link random walk on the original graph $G$ (see Fig \ref{frandwalks}B) now corresponds to exactly the same process as the usual vertex random walk on the original graph. This was not the case when dealing with undirected graphs, as  the sequence $\alpha - i - \beta - i -  \alpha$ is legitimate in terms of the link-node-link random walks on $G$, while it is not legitimate for a traditional vertex random walks, i.e. the single step $i-\beta-i$ is not allowed in the usual vertex walk process on $G$.  With directed graphs $G$ (assuming no self-loops) no edge can have the same source and target vertices so such a sequence never appears. In other words, the modularity for  line graphs $D(G)$, $E(G)$ and $F(G)$ defined for directed graphs are identical.  If this is advantageous one can always choose to represent an undirected graph as a directed graph to obtain these benefits.  However, it is not clear if these small differences between the random walks implicit in the construction of the line graphs will produce any significant differences in the analysis of a given network.

\section{Conclusion}

In this paper, we have extended our work on line graphs from unweighted \cite{EL09} to weighted graphs.  We have shown that this generalisation leads to the construction of line graphs which are both weighted and directed.  The goal of this simple and natural procedure is to move the focus from vertices to edges in the original graph for any graph based problem.

To illustrate this general principle we have used our weighted line graphs in the context of community detection.  The most popular schemes consist in partitioning the vertices of the graph, namely in assigning each vertex to a unique community.  Unfortunately, this approach is known to be inadequate in the many systems where vertices naturally belong to several communities.  This is the case of social networks for instance, where individuals (vertices) may be a member of several different communities characterised by different types of relationship, e.g.\ family ties, a shared hobby interest, or work connection.  An edge partition is particularity well adapted to such situations, as it naturally produces overlapping communities, while preserving the sound mathematical foundations of graph partitioning theory.  Our approach has the additional advantage to be easily implementable as the construction of a line graph is straightforward and the vertex partitioning of the line graph by any standard algorithm directly produces the optimal edge partition of the original graph.  The cost in terms of computer memory and time is roughly $O(\langle k^2\rangle / \langle k \rangle )$ (the ratio of edges in the line graph to the original graph), while the human cost in terms of code development is minimal\footnote{Codes to construct weighted line graphs and optimise modularity are freely available for download on the webpages \texttt{http://sites.google.com/site/linegraphs/} and
\texttt{http://sites.google.com/site/findcommunities/}.}.

\begin{acknowledgement}
R.L. acknowledges support from the UK EPSRC.
\end{acknowledgement}

\end{document}